\documentclass[prl,twocolumn,aps,showpacs,superscriptaddress]{revtex4}
\usepackage{graphics}
\usepackage{epsfig}
\usepackage{amsmath}
\usepackage{amssymb}
\usepackage{verbatim}
\usepackage{color}
\usepackage{hyperref}

\begin{document}
\newcommand{\beq}{\begin{equation}}
\newcommand{\eeq}{\end{equation}}

\title{Programmable Mechanical Metamaterials}

\author{Bastiaan Florijn}
\affiliation{Huygens-Kamerling Onnes Lab, Universiteit Leiden, Postbus
9504, 2300 RA Leiden, The Netherlands}
\author{Corentin Coulais}
\affiliation{Huygens-Kamerling Onnes Lab, Universiteit Leiden, Postbus
9504, 2300 RA Leiden, The Netherlands}
\author{Martin van Hecke}
\affiliation{Huygens-Kamerling Onnes Lab, Universiteit Leiden, Postbus
9504, 2300 RA Leiden, The Netherlands}

\date{\today}

\begin{abstract}
We create mechanical metamaterials whose response to uniaxial compression can be programmed by lateral confinement, allowing monotonic, non-monotonic and hysteretic behavior.
These functionalities arise from a broken rotational symmetry which causes highly nonlinear coupling of deformations along the two primary axes of these metamaterials.
We introduce a soft mechanism model  which captures
the programmable mechanics, and  outline a general design strategy for confined mechanical metamaterials. Finally, we show how inhomogeneous confinement can be explored to create multi stability and giant hysteresis.
\end{abstract}


\maketitle
Metamaterials derive their unusual properties from their structure, rather than from
their composition~\cite{Wegener_reviewRPP2008}. Important examples of mechanical metamaterials
are auxetic (negative Poisson's ratio) materials~\cite{Lakes_science1987}, materials
with vanishing shear modulus~\cite{Milton_JMPS1992,Kadic_APL2012,Buckmann_Natcomm2014}, materials with negative
compressibility~\cite{Lakes_Nature2001,Nicolaou_NATMAT2012}, singularly nonlinear
materials~\cite{Wyart_PRL2008,Gomez_PRL2012} and topological
metamaterials~\cite{Kane_NATPHYS2014,Chen_arxiv2014,Paulose_arxiv2014}.
Of particular recent interest are mechanical metamaterials whose functionality relies
on elastic instabilities, such as  quasi 2D slabs perforated with a square array of
holes~\cite{Mullin_PRL2007,Bertoldi_JMPS2008,Bertoldi_AdvM2010,Overvelde_AdvM2012,
Shim_SM2013}.
When compressed, these ``holey sheets'' undergo a buckling-like pattern transformation, which can be explored to
obtain switchable auxetics~\cite{Bertoldi_AdvM2010}, chiral and phononic
properties~\cite{Wang_IJSS2012,Wang_PRB2013,Wang_PRL2014}, and 3D ``buckley balls''~\cite{Shim_PNAS2012,Babaee_AdvM2013}.
An important limitation common to all these metamaterials is that each mechanical functionality requires a different structure.

\begin{figure}[t!]
\centering
\includegraphics[clip,width=.8\columnwidth]{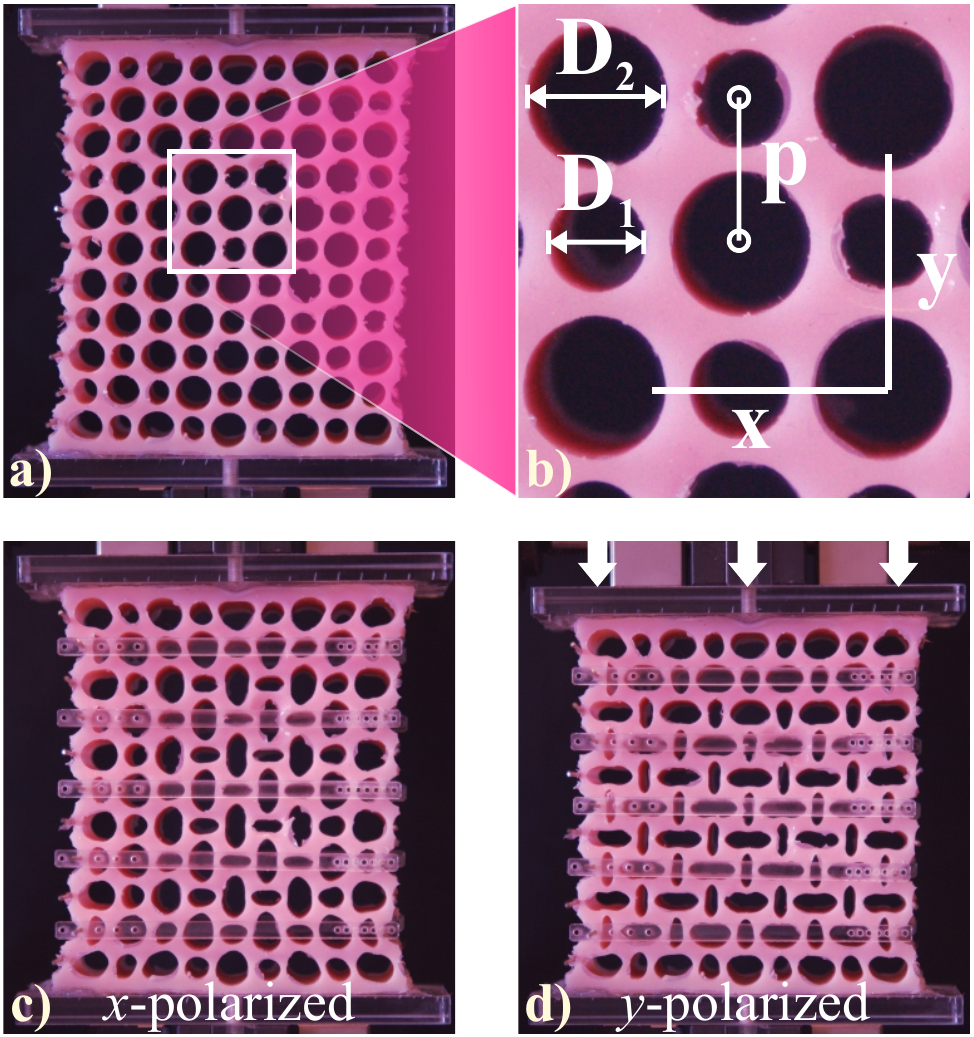}
\caption{(Color online) (a-b) A biholar sheet characterized
by  hole diameters $D_1=10$ mm, $D_2=7$ mm and pitch $p=10$ mm.
(c) Uniform confinement in the $x$-direction (by semi-transparent clamps) leads to
an $x$-polarization of the material.
(d) Sufficient $y$-compression of this $x$-confined material leads to
a switch of the polarization, and concomittant non-trivial mechanical behavior.}
\vspace{-0.5cm}
\label{fig:pic}
\end{figure}

Here we present a novel strategy to create {\em programmable} mechanical metamaterials, where the
response of a {\em single} structure is determined by confinement.
The core idea is illustrated in Fig.~\ref{fig:pic} for a {\em biholar} sheet, a
quasi-2D
elastic slab of material patterned by a regular array of large and small holes.
The difference in hole sizes breaks one of the 90$^\circ$ rotation symmetries
which is present for equal hole sizes. This causes a difference in the
polarization of the hole pattern, depending on whether
$x$-compression or $y$-compression is dominant --- see Fig.~\ref{fig:pic}c-d
 \cite{notepolarize}.

By constraining this metamaterial in the $x$-direction, and then
compressing it in the
$y$-direction, the material undergoes a pattern switch from a $x$-polarized to a $y$-polarized
state, as illustrated in Fig.~1c-d. Depending on the magnitude of the $x$-confinement, this pattern switch
can be either smooth or discontinuous, and the
force-displacement curves for the $y$-deformations can be tuned from monotonic
to non-monotonic (unstable) to hysteretic
--- all for a single biholar sheet.

We observe this tunability in experiments on systems of various sizes as well as in numerical simulations of the unit cell ---  this is a robust phenomenon.
We introduce a simple model that qualitatively captures all these different mechanical
behaviors, and which allows a precise study of the bifurcation scenario that underlies this phenomenology. Finally,
we show that the sensitive nature of the hysteretic switching between $x$ and $y$-polarized patterns can be explored in larger systems, where controlled inhomogeneities lead to multi-stability and giant hysteresis. We suggest that confinement is a general mechanism for a much larger class of programmable mechanical metamaterials, and outline how our model opens a pathway towards rational design of these materials.

\begin{figure}[t!]
\includegraphics[width=.99\columnwidth]{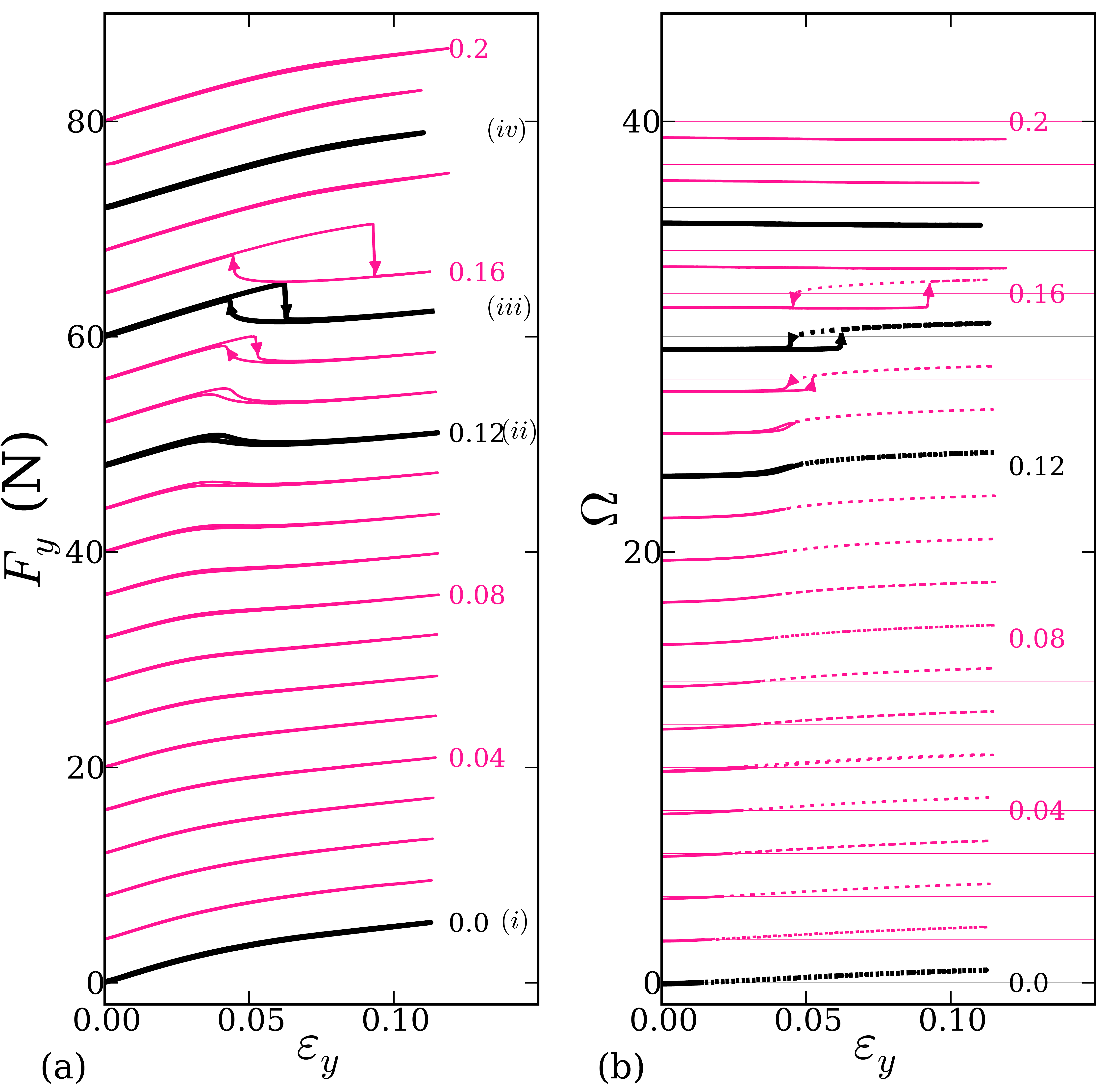}
\caption{(color online) Mechanical response of a $5\times5$ sheet with $D_1=10$~mm, $D_2=7$~mm ($\chi=0.3, t=0.15$)
for increasing confinement $\varepsilon_x$ \cite{note2}.
Curves are offset for clarity. (a) The force curves $F_y(\varepsilon_y)$ show
monotonic, non-monotonic and hysteretic behavior. The curves labeled {\em(i)-(iv)} correspond to four qualitatively different behaviors.
(b) Polarization of the
large center hole, where full (dashed) curves indicate negative (positive) values.
For movies of case {\em(i)-(iv)}, see \cite{supp}.
}
\label{fig:experiment}

\end{figure}
{\em Experimental Setup and Sample Preparation: }
We use quasi 2D elastic sheets of thickness 35 mm to avoid out-of-plane buckling. Their 2D structure is shown in Fig.~1.
The dimensionless numbers that characterize their geometry are the biholarity,  $\chi:= |D_1-D_2|/p$, and the dimensionless thickness of their most slender parts,
$t=1-(D1+D2)/2p$. To create these biholar sheets, we pour a two components
silicone elastomer (Zhermack Elite Double 8, Young's Modulus $E \simeq 220$ kPa,
Poisson's ratio $\nu\simeq 0.5$) in a mold which consists
of brass cylinders of diameter $D_1$ and $D_2$, precisely placed in the square
grid of two laser-cut
sidewalls with $p=10$ mm. We cut the sheets lateral sides and glue their flat bottom and top to Plexiglas plates, and then probe their
mechanical response in a uniaxial testing device
(Instron type 5965),
which controls the compressive strain $\varepsilon_y$ better than $10^{-5}$
and allows us to measure the compressive force $F_y$ with an accuracy $10^{-4}$~N.

To impose confinement we glue copper rods of diameter $1.2$~mm on the
sides of our material and use laser cut plastic clamps which have holes at distance $L_c$ to exert a fixed $x$-strain,
$\varepsilon_x:=1-L_c/Np$, where $N$ is the number of holes per row~\cite{note2}.
We image the deformations of the material with a CCD camera, resulting in a
spatial resolution of $0.03$~mm, and track the position and shape of the holes. This allows us to
determine the polarization of the holes, defined as $\Omega:= \pm (1-p_2/p_1)\cos2\phi$,
where $p_1$ ($p_2$) is the major (minor) axis and $\phi$ the angle
between the major axis and $x$-axis --- as the deformations of the small and big
holes are perpendicular, we define the sign of $\Omega$ such that $\Omega$ is
positive (negative) for $y$-polarization ($x$-polarization)~\cite{notepolarize}.


{\em Programmable Mechanical Response:} As we will see, inhomogeneities play an important role for large systems, so to probe the mechanics of asymptotically large
{\em homogeneous} materials, we start our exploration with the smallest experimentally feasible building block, a $5\times5$ biholar sheet.

In Fig.~\ref{fig:experiment} we present the force
$F_y(\varepsilon_y)$ and polarization $\Omega(\varepsilon_y)$ of the large center
hole for a range of values of $\varepsilon_x$.
We can distinguish four qualitatively different types of mechanical response.
{\em(i)} For $\varepsilon_x \lesssim 0.09$, the force increases monotonically with $\varepsilon_y$
and the polarization smoothly
grows from its initial negative value to positive values.
{\em(ii)} For $0.10 \lesssim \varepsilon_x \lesssim 0.13$, $F_y(\varepsilon_y)$ becomes
non-monotonic --- here the material has negative uniaxial compressibility. The increase in polarization gets focussed in the ''negative slope'' regime but remains monotonic.
{\em(iii)} For $0.14 \lesssim \varepsilon_x \lesssim0.16$,
$F_y(\varepsilon_y)$ exhibits a hysteretic transition with a corresponding hysteretic switch from $x$ to $y$ polarization. Note that in this case, the polarization
is non-monotonic and
initially {\em decreases} --- hence $y$-compression makes the center hole
initially more $x$-polarized.
{\em(iv)} For $\varepsilon_x \gtrsim 0.17$ the material becomes increasingly strongly $x$-polarized and does no longer switch to $y$-polarization and $F_y(\varepsilon_y)$ is smooth and monotonic. We
note that in this case,
additional experiments reveal that initial compression in the $y$ direction
followed by  $x$-confinement brings the
material to a strongly $y$ polarized state. Hence, for strong
biaxial confinement there are two stable states,
that are so different that uniaxial compression is not sufficient to make them switch.

We have explored this scenario for a number of different
biholar sheets, and find similar
behavior, provided that $\chi$ is large enough and $t$ is not too large~\cite{inprep}. Moreover, as we will discuss below,
such behavior is also found in larger systems. Hence, we conclude that confined
biholar sheets can be tuned to exhibit {\em four} different types of mechanical behavior.

\begin{figure}[t!]
\includegraphics[width=1\columnwidth,]{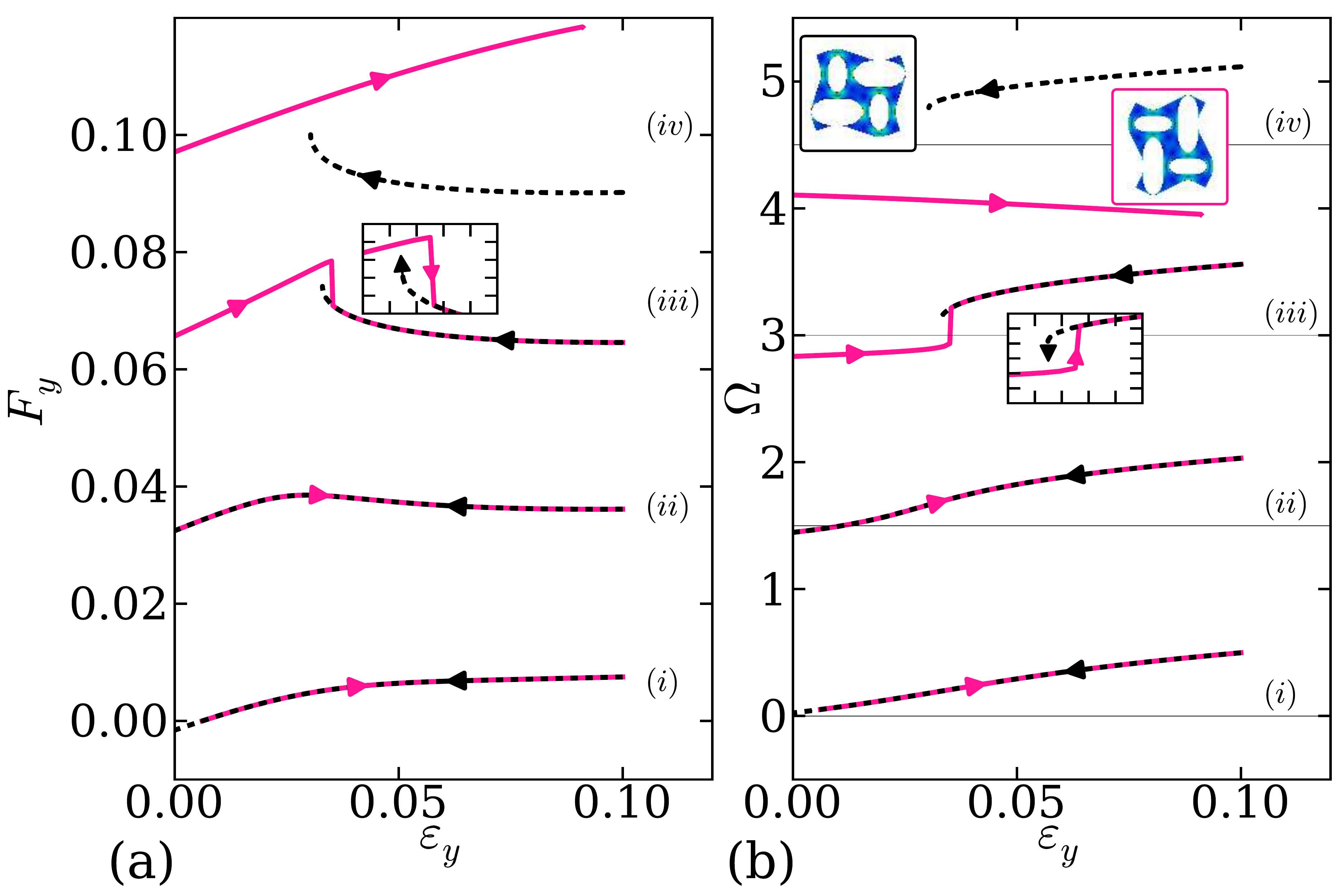}
\caption{(color online) Finite element simulations on a unit cell with periodic
boundary conditions and $\chi=0.3, t=0.15$. (a) Force $F_y$ and (b) polarization $\Omega$,
for {\em(i)} $\varepsilon_x=-1.0\times10^{-2}$, {\em(ii)}
$\varepsilon_x=2.8\times10^{-2}$,
{\em(iii)} $\varepsilon_x=3.4\times10^{-2}$, {\em(iv)} $\varepsilon_x=7.5\times10^{-2}$. Full (dashed) curves correspond to
protocol A (B).  For a movie of case {\em (iii)}, see \cite{supp}.}
 \label{fig:simu}
\end{figure}

{\em Numerics ---} To probe whether these phenomena are robust, we have performed 2D plane strain finite
element simulations (Abaqus) of a Neo-Hookean material on a
$2\times2$ unit cell using
periodic boundary conditions (Fig.~\ref{fig:simu}).
To probe hysteresis and bistability, we use two different protocols --- in protocol A
we first apply a confinement $\varepsilon_x$ and then a compression $\varepsilon_y$, whereas in protocol B
we first apply a large compression in the $y$-direction, then a confinement $\varepsilon_x$ and finally
decompress in the $y$-direction until we reach $\varepsilon_y$. As shown in Fig.~\ref{fig:simu}, these simulations exhibit
the four types of behavior {\em(i)-(iv)} observed in experiments. Moreover,
these simulations reveal that in case {\em(iv)}, the $x$ and $y$-polarized branches become disconnected,
consistent with our experimental data. The correspondence between experiments and simulations on systems with periodic boundaries show that our findings represent robust, bulk-type behavior, and suggest that biholar metamaterials of arbitrary size remain functional.

\begin{figure}[t!]
\includegraphics[width=1\columnwidth,]{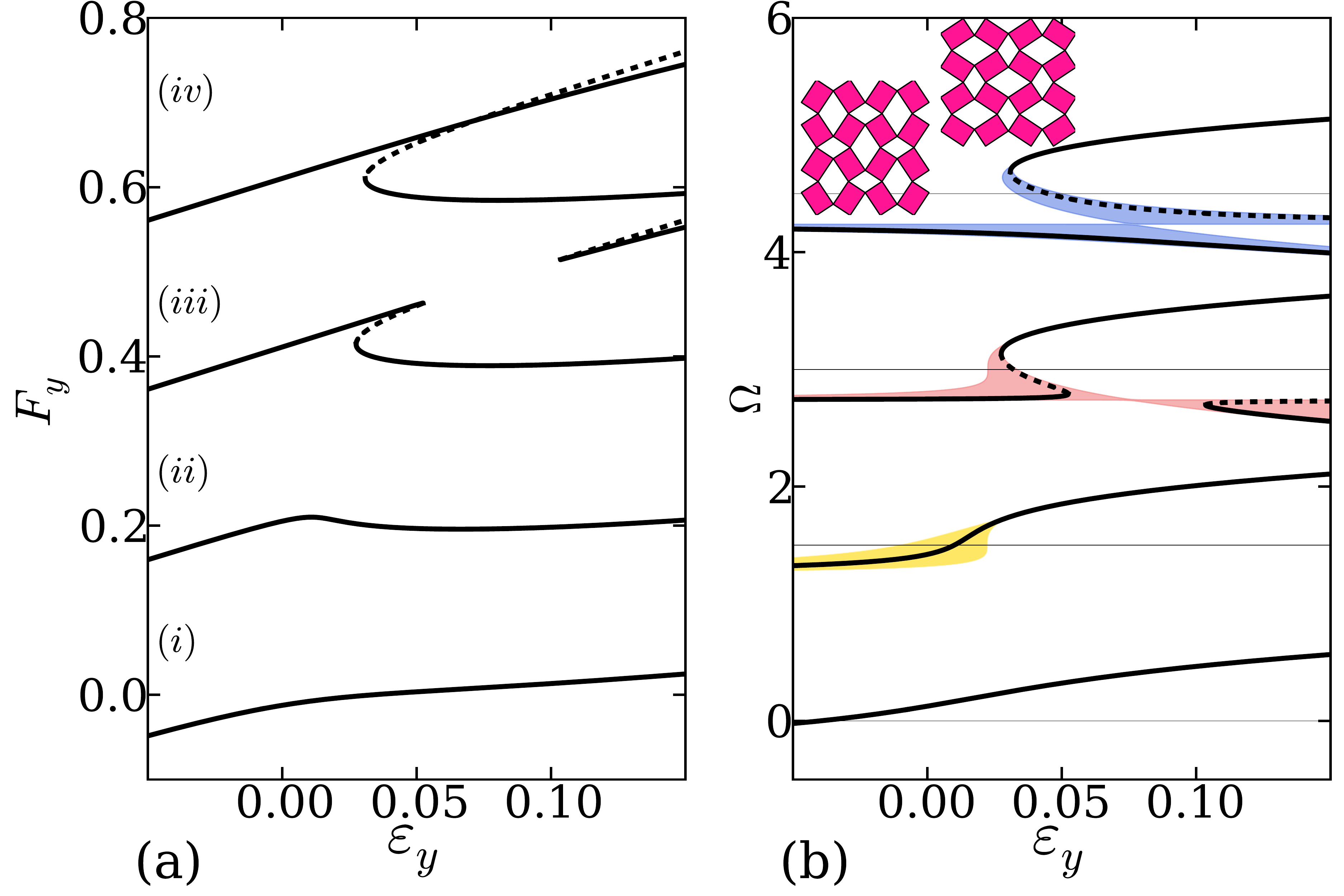}

\hspace{0.5cm}
\includegraphics[width=0.4\columnwidth,]{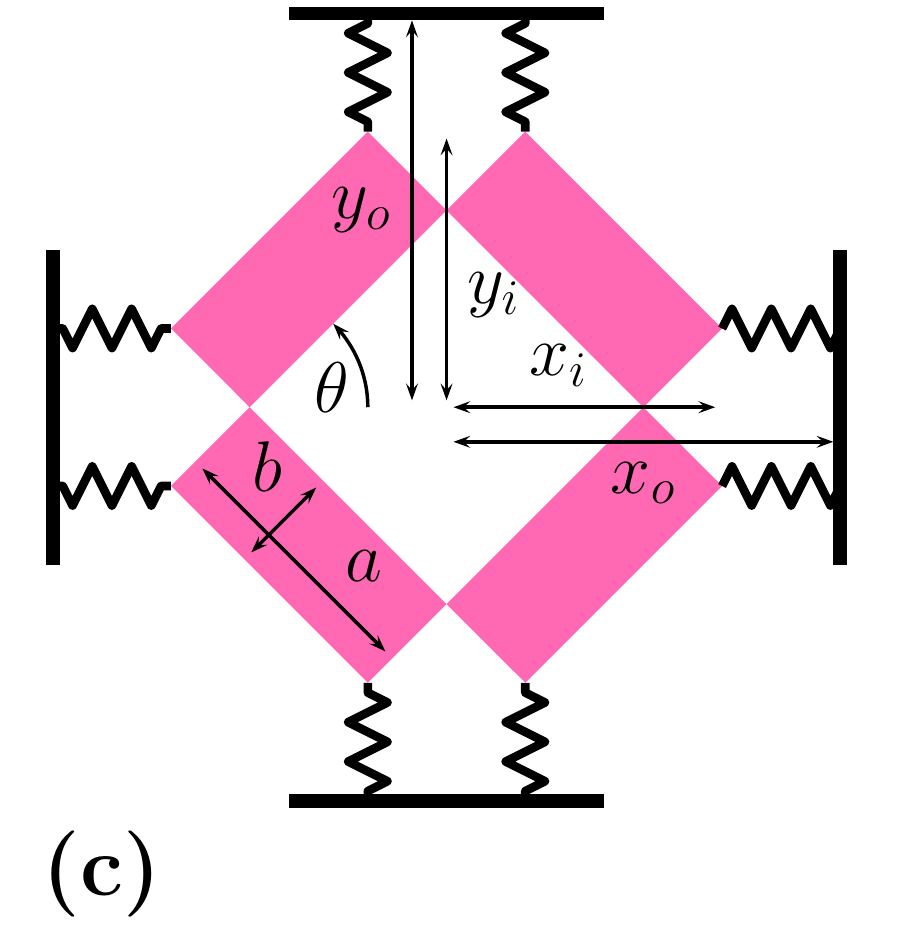}
\includegraphics[width=0.5\columnwidth,]{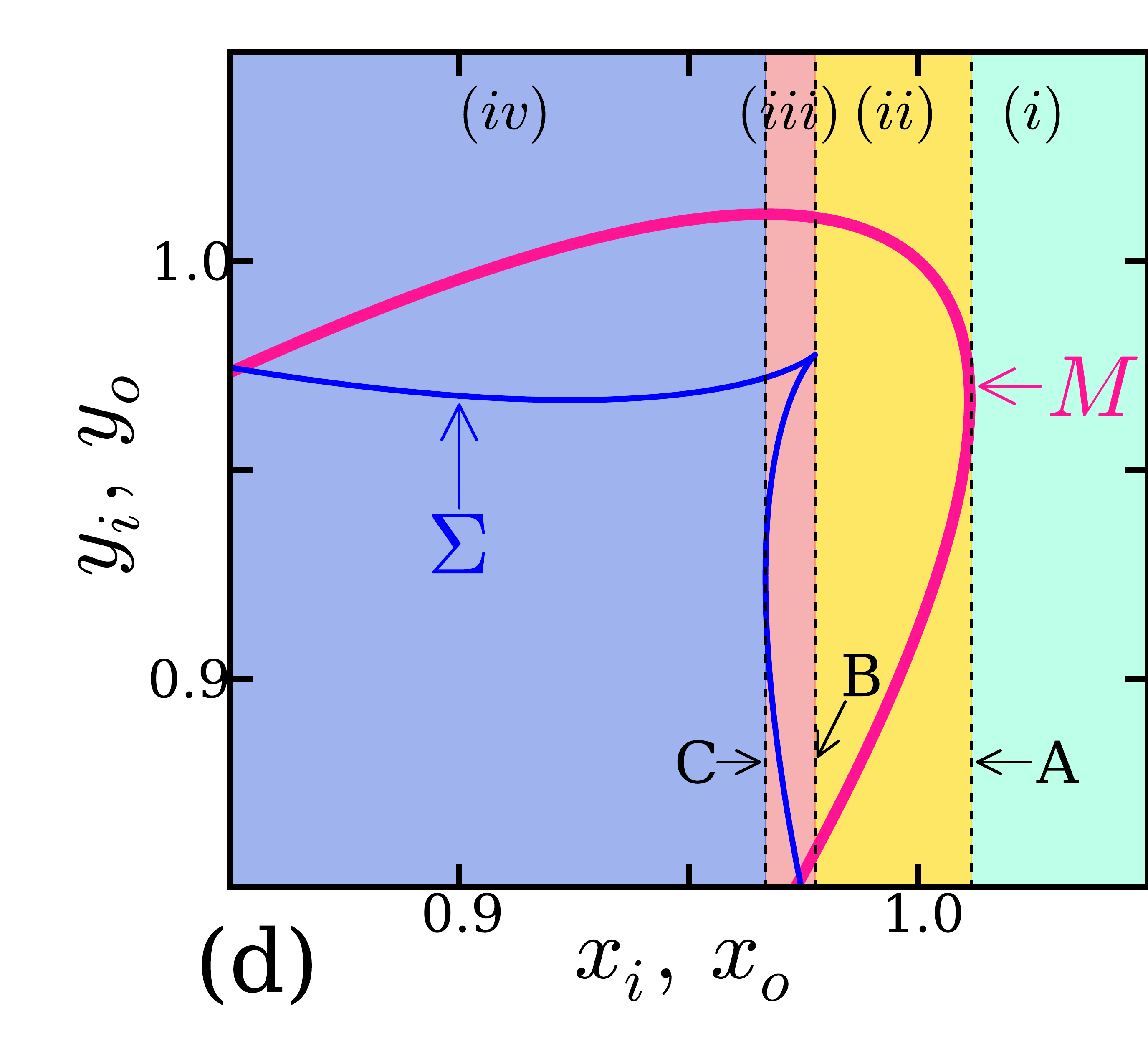}

\caption{(color online) (a-b) The force and polarization in our mechanical
model for $\chi=0.3$ ($a\approx0.81,
b\approx0.6$)
exhibit behaviors {\em(i)-(iv)} --- here full (dashed) curves correspond to stable (unstable) equilibria, and the insets show a $y$ and $x$-polarized state. The colored
regions in (b) are obtained by sweeping $x_0$ through the ranges {\em(i)-(iv)} and tracing out the corresponding polarization curves.
(c) Soft mechanism, where $\chi=2(a-b)/(a+b)$ and $a+b =\sqrt{2} $. (d) $M$-curve (thick solid pink), evolute $\Sigma$ (thin solid blue) and transitions {\em A - C} between domains {\em(i)-(iv)}. For movies illustrating case {\em(i)-(iv)} and {\em A - C} see \cite{supp}.
}
 \label{fig:model}
\end{figure}

{\em Soft Mechanism:} To qualitatively understand the mechanics of confined biholar sheets, we note that when $t
\rightarrow0$, the materials low energy deformations
are equivalent to that of a mechanism of rigid rectangles coupled by hinges
located at the necks of the ''beams''
(Fig.~\ref{fig:model}).
The state of this mechanism is described by a single degree of freedom, $\theta$, which determines the internal dimensions $x_i$ and $y_i$. To
model the storing of elastic energy, we couple this mechanism to outside walls
at spacing $x_o$ and $y_o$
via a set of linear springs with zero rest length and spring constant $1/2$
(Fig.~\ref{fig:model}c).
As Fig.~\ref{fig:model}a-b shows, such simple model qualitatively captures the full experimentally
and numerically observed scenario, when we identify the clamping $\varepsilon_x$ with
$1-x_o$ and compression $\varepsilon_y$ with $1-y_o$ (Fig.~\ref{fig:model}a-b).

\begin{figure*}[t!]
 \includegraphics[clip,height=0.44\columnwidth]{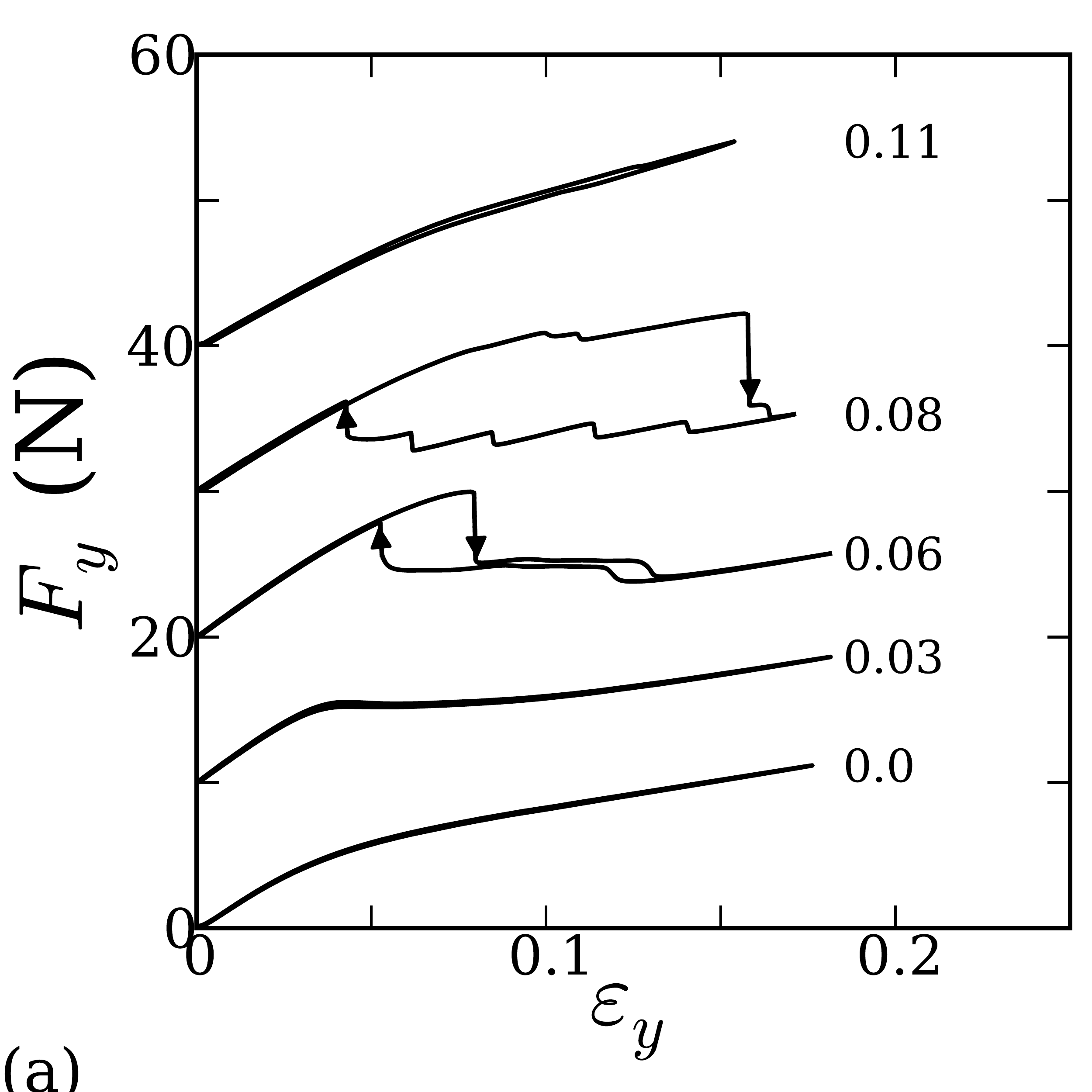} 
 \includegraphics[clip,height=0.44\columnwidth]{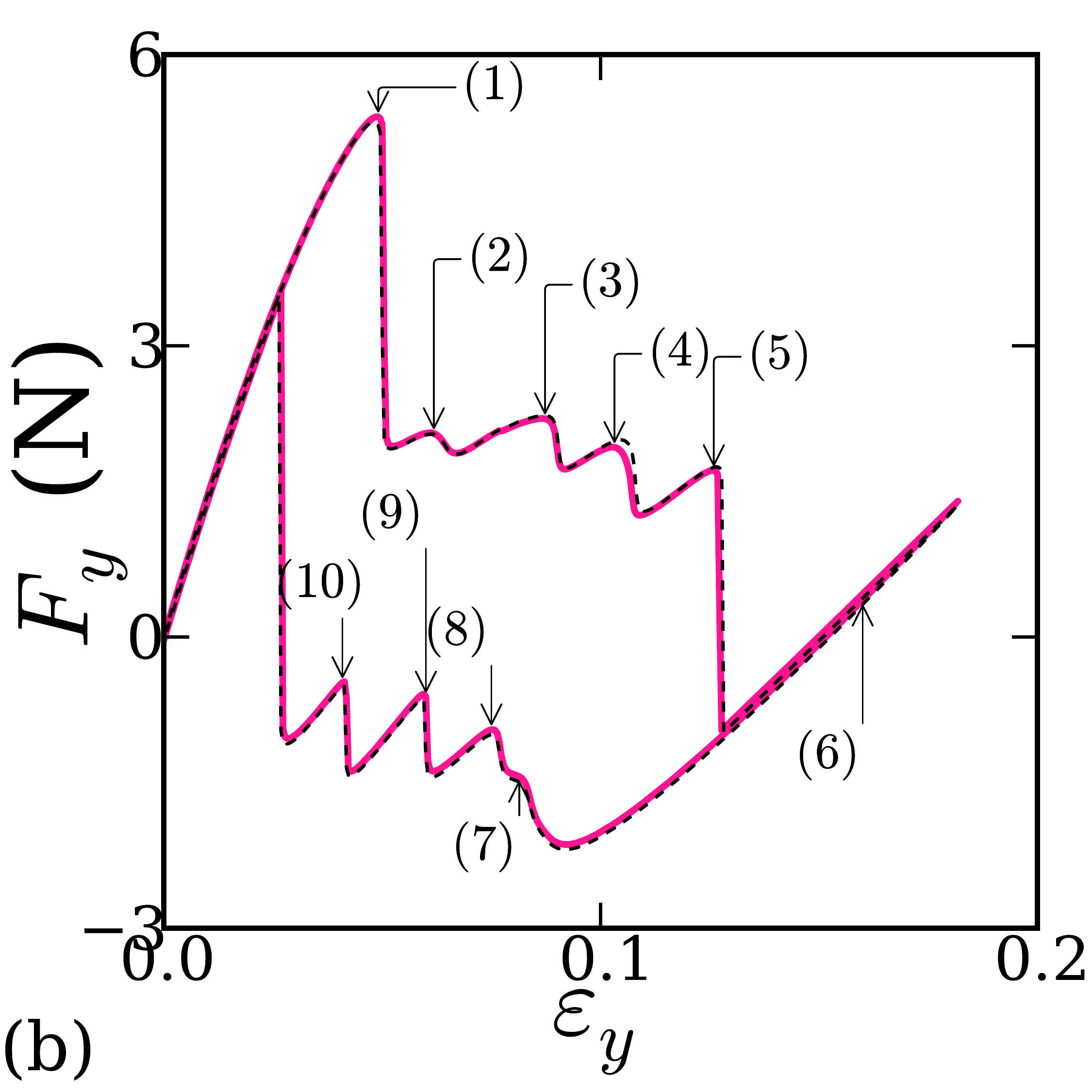} 
 \includegraphics[clip,height=0.43\columnwidth]{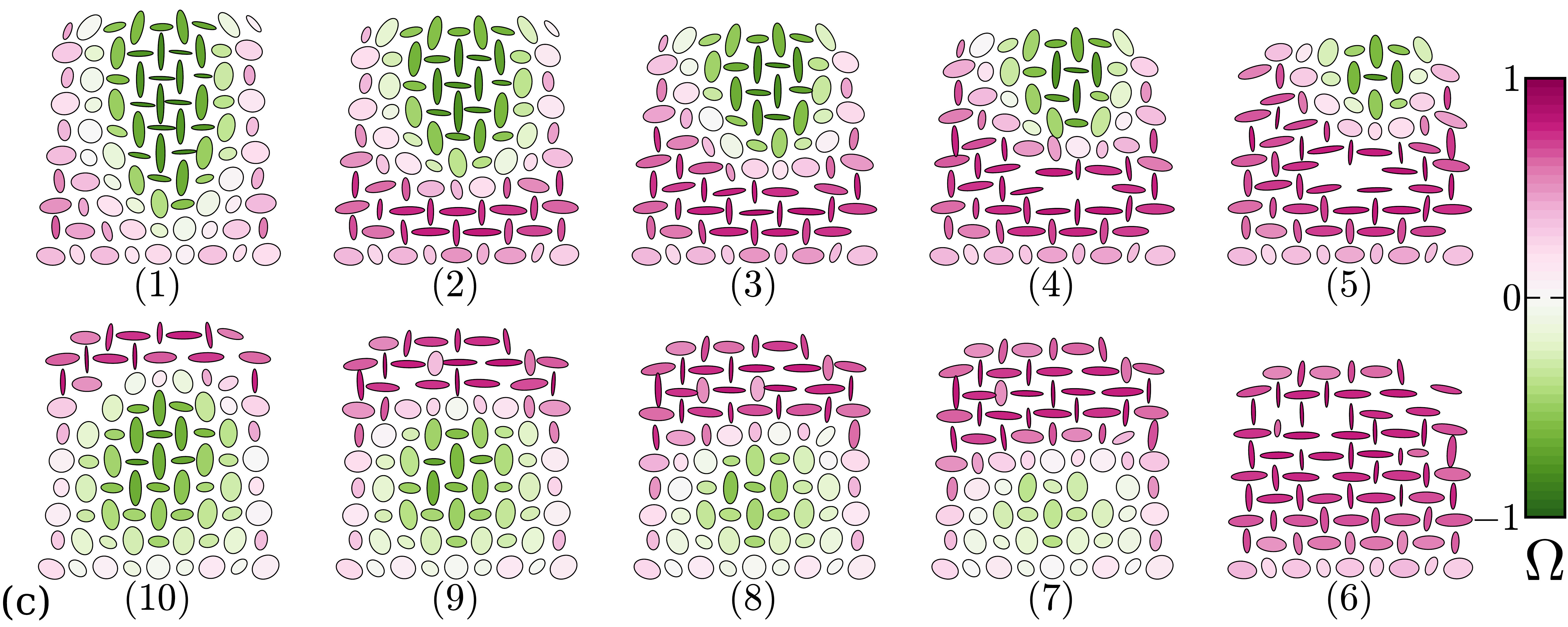} 
 \vspace{-.3cm}
\caption{{\bf (a)} Force-displacement curves for a $9\times11$ sheet
($D_1=10$ mm, $D_2=7$ mm,
$\chi=0.3, t=0.15$)
 with horizontal confinement $\varepsilon_x$
as indicated exhibits behavior {\em(i)-(iv)} as well as multistable behavior. Curves are offset for clarity. {\bf (b)} Force-displacement curve for an inhomogeneously confined $9\times11$ system, showing two giant hysteresis loops with multiple, perfectly reproducible polarization switching events. (c)
Snapshots of the state of our experimental metamaterial corresponding to (1)-(10), where color codes for polarization.}
\label{fig:multi}
\end{figure*}

A geometric interpretation of the various equilibria
and their bifurcations as $\varepsilon_x$ and $\varepsilon_y$ are varied provides much insight \cite{supp}.
As illustrated in Fig.~\ref{fig:model}d, the relation between the $x_i$ and $y_i$ can be
represented as a smooth curve, which we refer to as $M$ (for mechanism). For given $(x_o,y_o)$, the elastic energy $E$ equals
$(x_i-x_o)^2 + (y_i-y_o)^2$, so that equi-energy curves are circles of radius $E^{1/2}$. Stable (unstable) equilibria thus correspond to points on $M$ tangent to such circles, where $R_M$, the radius of curvature of $M$, is smaller (larger) than $E^{1/2}$. The experimental protocol varies $y_o$ at fixed $x_o$. Repeating this geometric construction while varying $y_o$ provides the corresponding stable and unstable equilibria, their elastic energies, and the force $F_y:=\partial_{y_o} E$.

We now explore this model to understand the transition {\em A} from monotonic to non-monotonic force curves, the transition {\em B} that leads to hysteresis, and the transition {\em C} where the differently polarized branches become separated.
In Fig.~\ref{fig:model}d we indicate the four trajectories corresponding to
the data shown in Fig.~\ref{fig:model}a-b, as well as three
trajectories labeled $A$, $B$ and $C$ that correspond to marginal curves which separate scenarios {\em (i)-(iv)}. To understand the transitions {\em B} and {\em C}, we
will consider the evolute $\Sigma$, the locus of all the centers of curvature of $M$
\cite{evolutewikipedia}. When $(x_o,y_o)$ crosses
$\Sigma$, saddle-node bifurcations
occur --- when $(x_o,y_o)$ crosses
$\Sigma$ in a non-generic manner, more complex bifurcations may arise ~\cite{guckenheimer1983nonlinear}.

Fig.~4d now gives a clear geometric interpretation of the three transitions:
{\em A:} Curve $A$ is tangent to $M$, so that here the
energy is purely quartic in $y_o$, and $\partial_y F_y =0$. Curve
$A$ thus separates {\em(i)} monotonic force curves at larger
$x_0$ from {\em(ii)} non-monotonic force curves for  smaller $x_0$.
{\em B:} Curve $B$ intersects the cusp of $ \Sigma$, leading to
a pair of saddle-node bifurcations which become separated for smaller $x_o$, and thus
spawn a hysteresis loop. Curve $B$ thus separates
case {\em(ii)} and {\em(iii)}.
{\em C:} Finally, curve $C$ is tangent to $\Sigma$, which
corresponds to a transcritical bifurcation where two solutions cross and
exchange stability. As a result, for smaller $x_o$, the differently polarized branches
decouple (Fig.~4b(iv)). Curve $C$ thus separates {\em(iii)} and {\em(iv)}, which can be seen as unfoldings of this transcritical bifurcation. For movies illustrating the geometrical construction for cases {\em(i)-(iv)} as well as {\em A-C}, see
\cite{supp}.

{\em Larger systems ---} To show that our observed mechanical functionality
can be experimentally realized in larger systems, we
show in Fig.~\ref{fig:multi}a examples of $F_y(\varepsilon_y)$ for
a $9\times 11$ sheet. We observe
the same four regimes as in Fig.~2 when the confinement is increased, illustrating the robustness
of these phenomena. The main difference with smaller systems is that
the hysteretic range is expanded and the force signal then has a complex structure
exhibiting several peaks. These correspond
to multiple switching events, due to inhomogeneities --- as hysteresis
corresponds to instability, even small inhomogeneities
are amplified.

We exploit this sensitivity to inhomogeneities to create multistable states.
We use five clamps of decreasing
length as function of row number, corresponding to strains
$\varepsilon_x=0.08,0.11,\dots,0.20$. As shown in Fig.~\ref{fig:multi}b, this results
in  a giant hysteresis loop with multiple peaks. We stress that Fig.~5b overlays two subsequent hysteretic loops, illustrating that
this complex behavior is well reproducible.
As shown in Fig.~\ref{fig:multi}c and in~\cite{supp},
each of these peaks corresponds to a
polarization switch of part of the material:
Under compression and decompression, a
polarization wave travels through our material.
We note that the spatial configurations in the downsweep of $\varepsilon_y$ are not the same as in the upsweep,
which can be understood from the observation that the most
confined part of the system shows the most hysteresis.

Hence, inhomogeneous confinement provides an avenue for the realization of complex multistable systems. Moreover, these
states with large hysteresis can also be seen as very effective dissipators of work, leading to novel strategies for mechanical damping \cite{Lakes_Nature2001,LakesPRL2001}.


{\em Outlook:} We have introduced a class of programmable
mechanical metamaterials whose functionality rests on two pillars:
First, confinement allows to store and
release elastic energy, crucial for complex mechanics. Second,
a broken symmetry leads to competition and coupling
between a secondary confinement and a primary strain.


The soft mechanism model suggests how to rationally design
mechanical metamaterials with confinement controlled response:
First, determine the bifurcation scenario when $\varepsilon_x$ is varied. Second, construct an evolute $\Sigma$ that is consistent with these sequence of bifurcations.
The $M$-curve can then explicitly be constructed as
the {\em involute} of $\Sigma$ \cite{wiki2}. Third, design a physical mechanism that
possesses this $M$-curve; in principle any $M$-curve is encodable in a mechanism~\cite{howtofoldit}. Finally, translate the rigid mechanism and hinges to a soft metamaterial with slender elements. Important work for the future is to explicitly
demonstrate the feasibility of this design strategy.

Finally, our work leads to several open questions, of which we highlight three.
First, for large, inhomogenous and multistable systems, how many distinct states
can be reached when more complex parameter sweeps are allowed? Second,
can we connect  the mechanics of biholar sheets
to the well-studied holey sheets with equal hole sizes, for which the mechanical response
is not fully understood
~\cite{Mullin_PRL2007,Bertoldi_JMPS2008,Bertoldi_AdvM2010,Overvelde_AdvM2012,
Shim_SM2013,Wang_IJSS2012,Wang_PRB2013}?
Third, can we use the strategies outlined here to  create fully 3D mechanical metamaterials?

\begin{acknowledgements}
We acknowledge discussions with K. Bertoldi,  K. Kamrin, P. Reis and S. Waitukaitis.
We thank H. Imthorn, J. van Driel and K. de Reus for exploratory experiments. BF, CC and MvH
acknowledge funding from the Netherlands Organization for Scientific Research NWO.
\end{acknowledgements}


\begin{thebibliography}{22}
\bibitem{notepolarize} Note that the major axis of the large ellipses is in the $x$ direction for $y$-polarized systems.
\bibitem{note2} We only clamp rows that end in small holes, to avoid subtle boundary effects~\cite{inprep}.
\end{thebibliography}

\end{document}